# Optical Tuning of Resistance Switching in Polycrystalline Gallium Phosphide Thin Films

*Fran Kurnia[†], Jan Seidel, Judy N. Hart, Nagarajan Valanoor\**

School of Materials Science and Engineering, UNSW Sydney, NSW 2052, Australia

The nanoscale resistive switching characteristics of gallium phosphide (GaP) thin films directly grown on Si are investigated as a function of incident light. Firstly, as-grown GaP films show a high $R_{ON}/R_{OFF}$ (~$10^4$), shown to arise from the formation of conductive channels along the grain boundaries. It is proposed that point defects (most likely Ga interstitials) and structural disorder at the grain boundaries provide the ideal environment to enable the filamentary switching process. Next, we explored if such defects can give rise to mid-gap states, and if so could they be activated by photonic excitation. Both first-principles calculations as well as UV-vis and photoluminescence spectroscopy strongly point to the possibility of mid-gap electronic states in the polycrystalline GaP film. Photoconductive atomic force microscopy (phAFM), a scanning probe technique, is used to image photocurrents generated as a function of incident photon energy (ranging from sub band-gap to above band-gap) on the GaP film surface. We observe photocurrents even for incident photon energies lower than the band-gap, consistent with the presence of mid-gap electronic states. Moreover the photocurrent magnitude is found to be directly proportional to the incident photon energy with a concomitant decrease in the filament resistance. This demonstrates GaP directly integrated on Si can be a promising photonic resistive switching materials system.

The rapid progress of non-volatile memory research has enabled us to process large amounts of data by creating ultra-high-density data storage.[1,2] In the last two decades, resistive random access memory (RRAM) has attracted significant attention because of its high density three-dimensional (3D) metal-insulator-semiconductor crossbar structure with fast switching speed and low power consumption.[3-5] RRAM is also capable of performing neuromorphic computing to mimic the human brain ability[6-8] and remains one of the leading embedded logic and memory technologies currently being considered for such applications. Concurrently, the 'More than Moore' ideology has motivated the development of integrated thin film heterostructures, in which the active device material can possess diverse functionalities and is amenable to being manipulated by multiple external parameters. For instance, photoactive materials for conversion of light into electrical signals now find use in various semiconductor and photovoltaic technologies.[9-12] Indeed, materials that demonstrate light-induced control of electrical conduction and *vice-versa* (*i.e.* optoelectronic and photonic materials) have rapidly risen to be one of the most intensely studied families of solid-state materials.[13-15]

Gallium phosphide (GaP) is one such technologically important optoelectronic material. Firstly, it has long been exploited as a key III-V semiconductor material for photonic applications.[16-18] With regard to current CMOS, the small lattice mismatch of GaP (∼0.37%) along with its thermal compatibility with Si provide strategic advantages. Not only does it enable the direct integration of GaP into current Si-based technologies[19] but also serves as a basic architecture for the growth of other III-V materials to exploit their valuable properties.[20] GaP has a band gap of 2.41 eV, which makes it attractive for many solar-powered functionalities *e.g.* tandem solar cells[20-22] and as a photoelectrode material in photoelectrochemical cells.[23,24] In addition, the high refractive index ($n_0 > 3$) of GaP can provide a strong optical confinement, which makes it useful for communication technologies.[25-27]

In this work, we explore the nanoscale resistive switching (RS) and photoconductive characteristics of a polycrystalline gallium phosphide (GaP) film directly grown on a Si substrate. It builds on our previous work, in which we systematically investigated the origins of RS in polycrystalline GaP films.[28] Here we investigate if the RS characteristic can be tuned by an incident photon flux. First, we postulate on the plausible microstructural origins of RS in polycrystalline GB, namely Ga interstitials within grain boundaries (GBs). Spatial position-dependent conducting atomic force microscopy (cAFM) in conjunction with cross-section transmission electron microscopy finds a strong correspondence between structural disorder at the GB and highly conductive paths. To gain an insight into the effect of such local structural disorder on the electronic properties (*i.e.* band gap, intermediate levels, shallow/deep traps, *etc.*) of GaP, density functional theory (DFT) calculations were carried out. The DFT results suggest that the presence of a Ga interstitial in the GaP crystal would introduce an intermediate energy state, which could facilitate the electronic transition within the band gap. Photoluminescence spectroscopy corroborates the DFT predictions where we find that the photoexcitation process can occur at sub-band gap energies. Photoconductive atomic force microscopy (phAFM) was conducted to probe the electrical response of the GaP film as a function of varying incident photon wavelength (*i.e.* energy) values. First, we find photocurrents are generated even for sub band-gap incident excitation implying the presence of active mid-gap electronic states. Secondly, the generated photocurrent and filament resistance showed a direct correlation (proportionality) with the incident photon energy confirming that the photoexcitation process can enhance the surface conductivity of the GaP thin film surfaces under light irradiation.

**Figure 1**a shows a scanning electron microscopy image of the grain structure of the GaP film deposited on a Si substrate. Conducting atomic force microscopy (cAFM) with a Pt-coated tip was used to spatially map the conductivity of the surface of the GaP film (see Experimental Method for details). Figure 1b shows the local slope of GaP topography, which has several grain boundary (GB) regions (upper panel) and the corresponding current map is shown in Figure 1b (lower panel). Compared with the other regions, the current at the GBs is found to be at least three orders of magnitude higher as shown in Figure S1, Supporting Information. The bipolar resistive switching cycles of the Pt-tip/GaP/Si structure are shown in Figure 1c. Due to its highly insulating nature, we initially need to form a soft dielectric breakdown in the GaP film by switching the sample from the OFF to ON -state, which is the SET process, with a positive DC voltage as indicated by the arrow. The current begins to increase at a bias of ~5.6 V, which corresponds to the SET voltage ($V_{SET}$). $V_{SET}$ is the activation voltage and it is strongly related to the formation of the soft dielectric breakdown, *i.e.* conductive nanofilaments.[5] In this regard, the higher the $V_{SET}$, the higher the power required to 'switch-on' the device and thus, $V_{SET}$ is one of the key parameters to evaluate the switching performance of the device.[5] When the bias sweeps back from positive to negative DC voltage, the current suddenly decreases, which is the RESET process; the device is then back to the OFF-state and this completes a switching cycle. Subsequent switching cycles produce a similar current level with a narrower hysteresis window as shown in Figure S2, Supporting Information. To prevent the switching filament from permanent dielectric breakdown during the switching cycles, a compliance current of 5 nA is applied. The switching cycle was repeated at least 30 times at different GB regions and consistently shows $R_{ON}/R_{OFF}$ of ~$10^4$.

The distribution of $V_{SET}$ values for our Pt-tip/GaP/Si structure is shown in Figure 1d (upper panel). It is interesting to note that the $V_{SET}$ values are rather scattered, with most values between 4.5 and 6.0 V but a small portion at 3.0 V, which could be explained as follows. The applied $V_{SET}$ is mostly used to form the conductive filaments at the GBs during the SET process; some GBs are likely to be more vulnerable for the filament formation because they have a higher defect concentration than other GBs. A non-uniform defect concentration distribution has also been found in other materials such as $TaO_x$,[3] $TiO_2$[29] and NiO,[30] which could be beneficial for triggering the soft dielectric breakdown in these materials. Furthermore, when we deposited a 50 × 50 $\mu m^2$ Pt pad as the top electrode on the GaP/Si and tested the switching cycles, the $V_{SET}$ distribution is much narrower compared with the Pt-tip/GaP/Si device structure as shown in Figure 1d (lower panel). Based on our previous results,[28] we attribute this to a larger coverage area of the Pt top electrode, which accommodates more conductive filaments underneath. In this case, the $V_{SET}$ values are mainly distributed between 1.0 and 2.0 V bias, which is on par with the most up-to-date non-oxide resistive switching devices.[31-35]

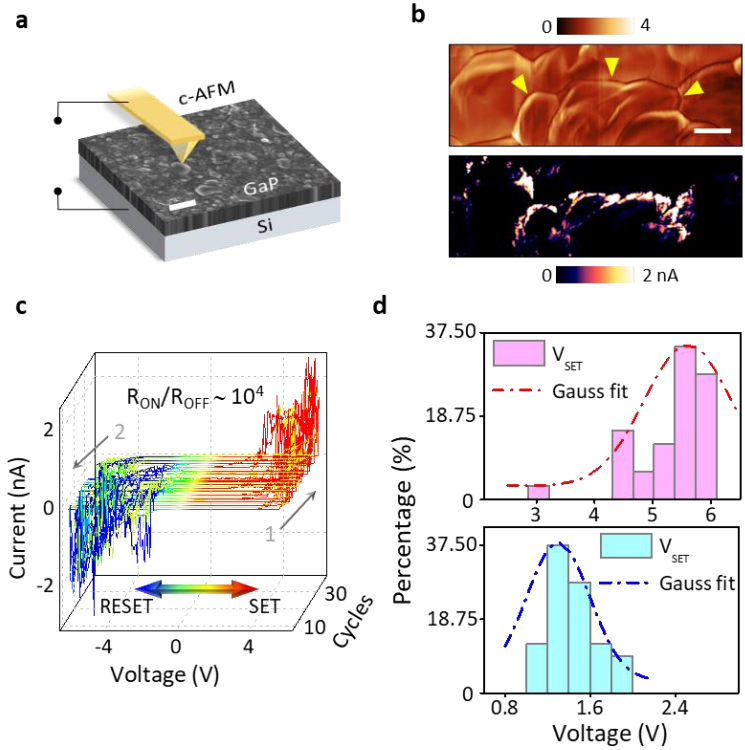

***Figure 1.*** *(a) Scanning electron microscopy (SEM) image of the GaP film grown on a Si substrate and a schematic illustration of the cAFM setup used to measure the resistive switching performance at the nanoscale. Scale bar is 500 nm. (b) Local slope (upper panel) and current map (lower panel) images of grain boundary areas in the GaP film. The yellow arrowheads indicate the location of the grain boundaries. Scale bar is 100 nm. (c) Repeated bipolar resistive switching of the GaP film; SET and RESET processes are indicated by the arrow at the bottom; the grey arrows labelled 1 and 2 indicate the direction of the applied voltage. (d) The $V_{SET}$ distribution of the Pt-tip/GaP/Si (upper panel) and Pt/GaP/Si (lower panel) devices.*

In materials such as GaP, the dynamic filament growth plays an important role as often it is the filamentary process that drives the RS properties and hence device operation and optimization. **Figure 2** illustrates the filament formation and rupture under application of an external bias for the GaP system of interest. Previously, x-ray photoelectron spectroscopy (XPS) measurements on our GaP films confirmed the presence of peaks that correspond to metallic Ga.[28] We suggested these Ga-rich regions were localized to the grain boundaries and hence formed conductive channels in the ON-state, but at the time did not have any direct evidence. The cAFM images now shown in Figure 1b confirm this hypothesis, *i.e*, there are regions of high conductivity that correspond to the grain boundaries. A strong possibility (based on the combination of our prior XPS and present cAFM data) is that the grain boundaries could host an array of point defects, such as Ga interstitials.

Thus the proposed mechanism of RS behavior is that, initially, the dielectric GaP layer is in the OFF-state (Figure 2a), in which there are Ga interstitial defects randomly distributed throughout the crystal structure. When a positive bias with respect to the bottom electrode (*i.e.* Si) is applied *via* the cAFM tip, the current abruptly increases at $V_{SET}$ (see Figure 1c) as a consequence of complete formation of conductive filaments (SET process); the device is then in the ON-state (Figure 2b). During the SET process, the Ga defects, which were randomly distributed in the crystal, diffuse and become localized at the GB regions to create conducting channels through the GaP layer. Cross-section TEM analysis (Figure S3, Supporting Information) indeed confirms significant structural disorder in the grain boundaries of the GaP film implying that localization of such interstitials and other point defects is definitely possible. In addition to the complete filaments that are formed, there would also be a few incomplete filaments owing to the different structures of the GBs. Unlike other metal-insulator-semiconductor devices, where the use of an active electrode material (*i.e.* Ag) is required to inject material into the dielectric layer to create a conducting channel,[5] our GaP does not require an active electrode for the conductive filament formation.

Having thus demonstrated direct evidence of RS driven by localization of point defects at grain boundaries, we next pose the question – what if such defects (*i.e.* Ga interstitials) lead to mid-gap states in GaP and, if so, can they be excited by light?

This is schematically depicted in Figures 2c-d. Note that the band gap ($E_{gap}$) of 2.41 eV[20] for a perfect stoichiometric GaP layer is high enough to prohibit electrons in the valence band from being excited to the conduction band under ambient conditions, leading to the insulating properties of the material as illustrated in Figure 2c. However we have already shown that II-VI/III-V semiconductor films (*e.g.* ZnS,[36] GaP,[28] *etc.*) deposited by pulsed laser deposition contain defects, attributed to excess cations. The electronic properties of such thin films are then strongly influenced by the presence of ionic/electronic defects, such as vacancies and interstitial atoms.[36,37] Here these would be defects associated with excess Ga in the as-deposited GaP layer.[36] Following on from the model of Szot *et al.*[38] for oxides, it is plausible that the these defects introduce a mid-gap energy state (hereafter called an *intermediate band*, IB) between the host valence band and conduction band (Figure 2d). This would influence the overall electronic/photoelectric properties of GaP.

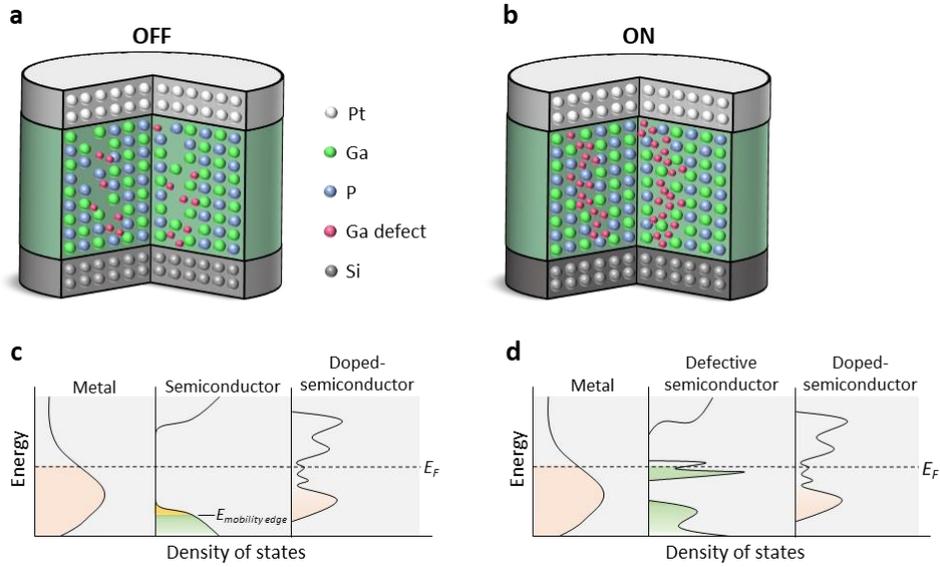

***Figure 2.*** *Schematic illustration of (a) the switching filament rupture (OFF-state) and (b) formation (ON-state) along the grain boundaries (GBs) of the GaP film. The proposed densities of electronic states of a (c) defect-free GaP structure, in which the electrons located at the mobility edge are inactive due to the large band gap and (d) defective GaP, in which the presence of a mid-gap energy state due to the crystal impurities leads to its conducting properties.*

Therefore, we next performed density functional theory (DFT) calculations of both pure GaP and GaP containing a Ga interstitial, to investigate the impact of such Ga point defects on the electronic structure, as shown in **Figure 3**. Figure 3a shows the zinc blende GaP lattice structure (space group $F\bar{4}3m$), in which each atom is bonded in a tetrahedral coordination environment (see inset). For this pure crystalline GaP, an indirect $E_{\text{gap}}$ of 2.48 eV is calculated as shown in Figure 3b, which is slightly larger than the experimentally reported value of 2.41 eV.[20] The valence band is formed by a bonding state of overlapping Ga $4s$ and P $2p$ atomic orbitals, while the conduction band is formed by the corresponding antibonding combination (Figure S4,

Supporting Information). When a single interstitial Ga defect is introduced into the GaP crystal lattice, as shown in Figure 3c, a fully occupied mid-gap state is formed above the valence band of pure GaP as indicated in Figure 3d. This indicates that point defects in GaP can indeed produce mid-gap states. In synthesized materials, it can be expected that a variety of point defects may form, producing mid-gap states at various energies. These mid-gap energy states can assist to excite electrons from the valence band to the conduction band at lower energies than for pure GaP, resulting in increased conductivity of the material.[36,37]

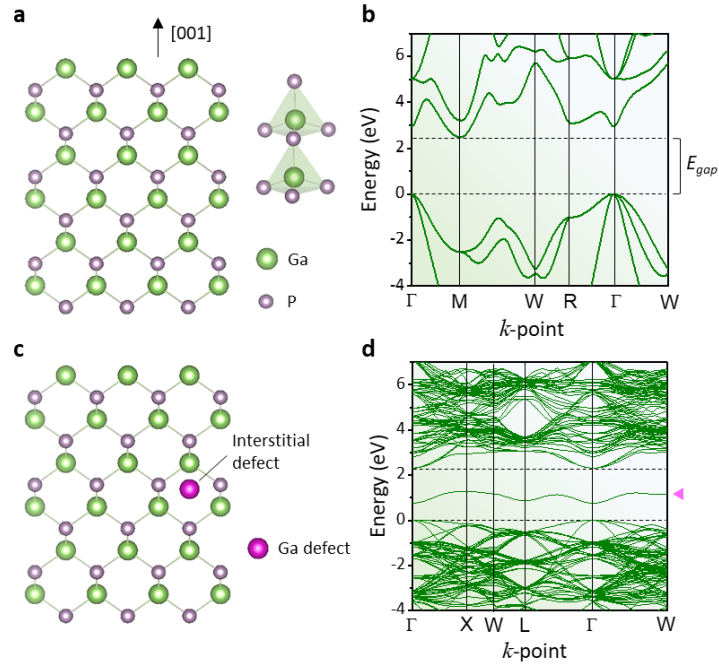

**Figure 3.** (a) The zinc blende structure adopted by GaP; two Ga-centred tetrahedral units of the crystal structure are shown on the right. (b) Electronic band structure of GaP calculated by DFT. (c) Ga-rich GaP formed by introducing an interstitial Ga ion into the pure GaP crystal lattice. (d) Electronic band structure of Ga-rich GaP. The magenta arrowhead indicates the mid-gap energy state introduced within the band gap of GaP by the interstitial Ga. The edge of the valence band is positioned at 0 eV.

The next step was to check for evidence that such mid-gap states do indeed exist in our films. For this, a UV-vis spectrum was collected at wavelengths from 300 to 800 nm and photoluminescence spectroscopy was acquired, (data shown in supplementary S5). Two (sub-band gap) photoemission peaks at longer wavelengths (583 nm and 767 nm corresponding to 2.13 eV and 1.62 eV, respectively) are observed. This suggests the presence of mid-gap states as the band-gap of GaP is ~2.41 eV.[20] This also implies electronic transitions through the defect states are more prominent in our GaP film than the valence-to-conduction band transition.

If such mid-gap energy states indeed exist, their influence on the electrical conductivity should be readily deduced by shining an incident photon flux with the appropriate wavelengths on the GaP surface. Thus photoconductive atomic force microscopy (phAFM) was employed to directly image photocurrent generation on the GaP thin film surface due to an incident photon flux. phAFM is a powerful tool to directly map photoinduced currents flowing in semiconductors under different photon wavelengths.[39] Since GaP has negligible two-photon-absorption (TPA),[16-18,40] it is well-suited for examination by phAFM to visualize the photocurrent generated by absorption of photons at specific energies.[39,41] **Figure 4**a shows a schematic illustration of our phAFM experiment (further details are provided in the Experimental Methods, Supporting Information). The topography map of the GaP film shows a granular structure with ~0.31 μm grain height (on average) and a few GBs as indicated by the line profile shown in Figure 4b. The film surface was illuminated at photon energies between 2.95 eV ($\lambda_{photon}$ = 420 nm) and 2.17 eV (570 nm) to induce the photoexcitation process. The corresponding photocurrent maps are shown in Figure 4c. A photocurrent is produced even with sub-band gap energy illumination; this confirms both the presence of mid-gap states and their role in the photoexcitation process. At sub-band gap energies, the current level is rather low (~2 pA at 2.17 eV photon energy), but when the photon energy is increased, the current level, especially at the GB regions, appears to be increasing. An overall current enhancement of a factor of ~40 could be observed in the film surface when the sample was illuminated by 2.95 eV photons compared with 2.17 eV, indicating that our GaP film is photoresponsive.

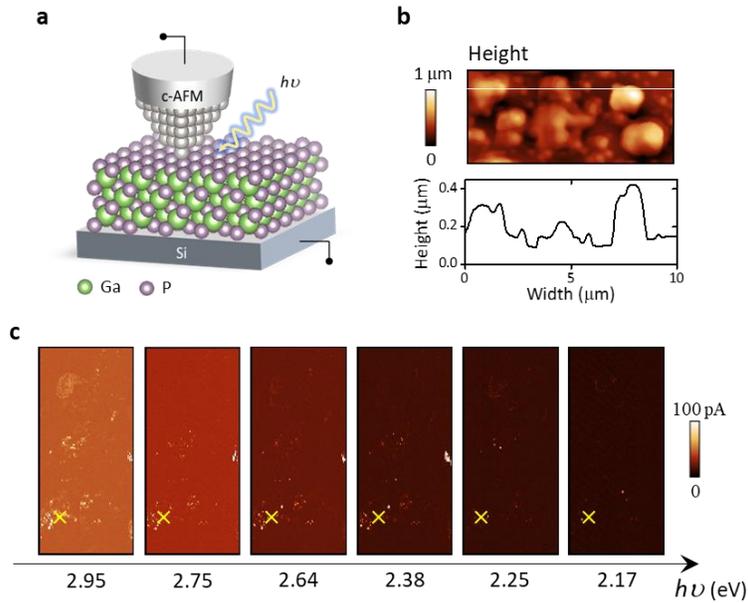

***Figure 4.*** *(a) Schematic illustration of the photoconductive atomic force microscopy (phAFM) measurement. (b) Surface topography of the GaP film imaged by phAFM. The lower panel shows a line profile at the position indicated by the white line in the upper panel, with sudden changes in height in the profile corresponding to the presence of grain boundaries. (c) Photocurrent maps of the area in (b) for different incident photon energies (hυ). The yellow cross symbol indicates the point where the current-voltage sweep is applied.*

Current-voltage sweeps were obtained from a representative GB region for each specific photon energy as shown in **Figure 5**a. It is interesting to note that the switching current changes significantly as a function of the photon energy. From the current-voltage curves, the $R_{ON}/R_{OFF}$ and filament resistance ($R_0$) can be extracted. The filament resistance is the initial resistance when the filament begins to form and thus it can be calculated from the generated current at low bias voltage ($V_{low}/I$).[28] Figure 5b shows the change of the $R_{ON}/R_{OFF}$ and $R_0$ as a function of photon energy. It can be seen that the $R_{ON}/R_{OFF}$ is rather scattered, while interestingly, the $R_0$ shows a decrease with increasing photon energy. This suggests that the light absorbed by the GaP material tends to induce carriers to improve its conductivity. Thus, light illumination can provide more excited charge carriers to improve the generated photocurrent in GaP films as shown in Figure 5c; the presence of the IB allows this to occur at lower photon energies than for defect-free GaP.

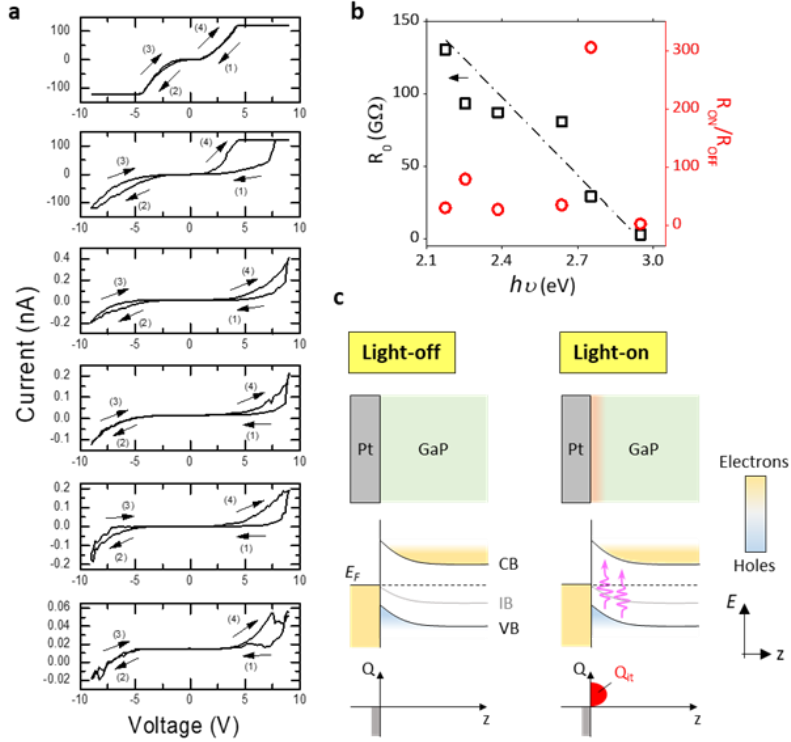

***Figure 5.*** *(a) Current-voltage (I-V) curves measured under different wavelengths of light irradiation. The incident photon energy for each I-V loop from top to bottom panel is 2.95, 2.75, 2.64, 2.38, 2.25, and 2.17 eV. (b) Filament resistance ($R_0$) and $R_{ON}/R_{OFF}$ plot as a function of photon energy. The dash-dotted line is a linear fit of the $R_0$ data and serve to guide the eye. (c) The proposed photoexcitation mechanism at the Pt/GaP interface without light illumination (left panel) and under the light illumination (right panel).*

In summary, we have investigated the resistive switching characteristics and photoconductive properties of GaP directly grown on a Si substrate. A repeatable resistive switching was observed at the grain boundary regions of the GaP film with $R_{ON}/R_{OFF}$ ~$10^4$. This resistive switching is attributed to the formation of filaments, likely to be formed by excess Ga that localizes in the grain boundaries. First-principles calculations predict that such Ga interstitials are capable of forming intermediate band states, and the presence of such states was confirmed by UV-vis and PL spectroscopy. These states can effectively reduce the energy required for photoconduction. phAFM measurements show (i) a systematic increase in surface conduction as a function of increasing photon energy and (ii) photoactivity at sub-band gap energies, confirming the role of mid-gap states. These results show that a GaP film directly grown on Si is a promising candidate material for non-volatile resistive switching memory and nanophotonic applications.

ASSOCIATED CONTENT

Supporting Information files are available free of charge.

Experimental and simulation method (PDF); Figure S1: Current profile across grain boundaries; Figure S2: Subsequent I-V curves; Figure S3: Cross-section TEM image of GaP film; Figure S4: Density of states of GaP and defective GaP; Figure S5: UV-vis and photoluminescence spectroscopy data (PDF)

AUTHOR INFORMATION

**Corresponding Author**

*nagarajan@unsw.edu.au

**Present Addresses**

†Center for Nanostructured Material, School of Mathematics and Physics, Queens University Belfast, Belfast BT7 1NN, Northern Ireland, United Kingdom

**Author Contributions**

The manuscript was written through contributions of all authors. All authors have given approval to the final version of the manuscript.

ACKNOWLEDGMENT

The DFT calculation was undertaken with the assistance of computational resources provided by the Australian Government through the National Computational Infrastructure (NCI) under the National Computational Merit Allocation Scheme. The authors thank Nastaran Faraji (University of New South Wales) for the assistance in photoconductive atomic force microscopy measurement.

A current enhancement up to 40 times higher can be observed under light illumination of the film, which benefits from the formation of an intermediate band in the Ga-rich film.

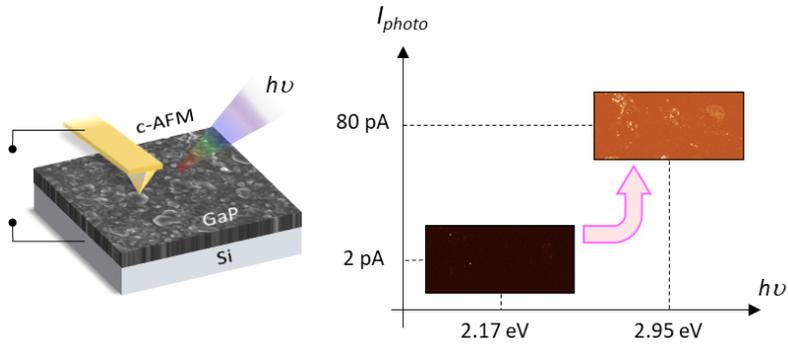